\documentclass[aps,twocolumn,superscriptaddress,showpacs,amssymb,prl,floatfix]{revtex4-1}
\usepackage{epsfig}
\usepackage{graphicx}
\usepackage[dvipsnames,usenames]{color}
\usepackage{color}
\usepackage{amsmath}
\usepackage{epstopdf}

\usepackage[pdftex]{hyperref}

\hypersetup{colorlinks=true,linkcolor=blue,citecolor=blue,urlcolor=blue}

\emergencystretch=\maxdimen
\begin{document}
\title{Impurity-Induced Antiferromagnetic Domains in the Periodic 
Anderson Model}

\author{A.~Benali}
\email{ali.benali@fst.rnu.tn}

\affiliation{Department of Physics, Faculty of Sciences of Tunis,
University of Tunis El-Manar, Tunis 2092, Tunisia}

\affiliation{Department of Physics, One Shields Ave.,
University of California, Davis, California 95616, USA}

\author{Z.J.~Bai}
\email{bai@cs.ucdavis.edu}

\affiliation{Department of Computer Science, One Shields Ave.,
University of California, Davis, California 95616, USA}

\author{N.J.~Curro}
\email{curro@physics.ucdavis.edu}

\author{R.T.~Scalettar}
\email{scalettar@physics.ucdavis.edu}

\affiliation{Department of Physics, One Shields Ave.,
University of California, Davis, California 95616, USA}

\begin{abstract}
A central feature of the Periodic Anderson Model is the competition
between antiferromagnetism, mediated by the
Ruderman-Kittel-Kasuya-Yosida interaction at small conduction
electron-local electron hybridization $V$, and singlet formation at
large $V$.  At zero temperature, and in dimension $d>1$, these two
phases are separated by a quantum critical point $V_c$.  We use Quantum
Monte Carlo simulations to explore the effect of impurities which have a
local hybridization $V_{*} <  V_c$ in the AF regime which are embedded
in a bulk singlet phase with $V > V_c$.  We measure the suppression of
singlet correlations and the antiferromagnetic correlations which form
around the impurity, as well as the size of the resulting domain.  Our
calculations agree qualitatively with NMR measurements
in CeCoIn$_{5-x}$Cd$_x$.  
\end{abstract}

\pacs{71.10.Fd, 71.30.+h, 02.70.Uu}
\maketitle

\section{Introduction}

The interplay of disorder and interactions results in intriguing
metal-insulator, superconductor-insulator, and magnetic order-disorder
transitions.\cite{lee85,belitz94,ulmke95,ulmke97,kravchenko04,DobrosavljevicDisorderReview2005,goldman10}  
A particularly rich set of questions arises when randomness is introduced
to a system which is already close to a critical point, for example
through tuning the electron density or interaction strength.  Here the
effects of impurities might be expected to be especially large, since
the system is already poised on the brink of two distinct phases.

One example of this situation is provided by the replacement of Cu by
nonmagnetic atoms in cuprate superconductors where neutron scattering
studies of Zn-doped La$_{2-x}$Sr$_x$CuO$_4$ reveal the emergence of
magnetic scattering peaks and the emergence of a novel  static spin
state within the spin gap.\cite{kimura03} Insight into this phenomenon
was provided by Hartree-Fock calculations on the single band Hubbard
Hamiltonian which examined the effect of local chemical potential shifts
on the striped phase of coexisting d-wave superconductivity and AF
order.\cite{andersen07} Local AF order was found to nucleate about the
impurities above a critical threshold for the on-site interaction $U$.
This local phase transition occurs at a different $U_c$ for each
impurity.  Further theoretical work investigating the nature of local
defects in quantum critical metallic systems indicates that large
droplets are formed with a suppression of quantum
tunneling.\cite{DefectsQCmetals} A remarkable feature of both theory and
experiment is the large extent of AF order induced by small impurity
concentration.  The cuprate problem is made even more complex by the
large degree of inhomogeneity, eg in the superconducting gap, charge and
magnetic stripes, etc, that are present even without Zn
doping.\cite{phillips03,keimer15}


Recent experiments on the replacement of In by Cd in CeCoIn$_5$ examine
closely related phenomena in heavy fermions
materials.\cite{ParkDropletsNature2013} In this case, the parent compound
CeCoIn$_5$  is already a quantum critical superconductor without the
necessity of resorting to pressure or chemical doping, as is the case
for La$_{2-x}$Sr$_x$CuO$_4$.  In doped CeCoIn$_5$, it was found that AF
islands develop about the Cd impurities, and ultimately coalesce into a
magnetically ordered, but very heterogeneous, phase.\cite{Urbano2007}
The inhomogeneous response of the electronic system clearly demonstrates
that doping does not necessitate a quantum critical response, as
observed in other heavy fermion
systems.\cite{CeCu6AuNeutronsPRL98,Friedemann2009} Related issues
concerning anomalous non-Fermi liquid phases which intervene the Kondo
singlet to AF transition in doped $f$ electron alloys have also been
explored.\cite{proceedings96,castroneto98,vojta10}

In analogy with numerical work on the single band Hubbard Hamiltonian
appropriate to modeling the cuprates, it is natural to consider a two
band Periodic Anderson Model (PAM)\cite{anderson61} to understand these heavy fermion
experiments.\cite{andersen07} One of the basic features of the PAM (when
there is one electron per site) is the competition between an AF phase
when the hybridization $V$ of the conduction and localized electrons is
small, and a singlet phase when $V$ is large.  Since there is no direct
coupling between the local moments, their ordering arises through the
Ruderman-Kittel-Kasuya-Yosida (RKKY)
mechanism\cite{ruderman54,kasuya56,yosida57} in which polarization of
the spin of the conduction electrons mediates an indirect interaction.
In the absence of randomness, Quantum Monte Carlo (QMC) has quantified
this AF-singlet transition, both in the itinerant electron
PAM,\cite{vekic95}, and also in quantum spin Hamiltonians which are the
strong coupling limits of the PAM, like the bilayer Heisenberg model
where the interlayer exchange $J_{\perp}$ is
varied.\cite{sandvik94,sandvik06} The position of the quantum critical
point (QCP) in the uniform
system is also known in the case when one of the layers has no
intra-layer exchange, a geometry which is similar to that of the PAM
where the $f$ electrons are localized and have $t_{\rm
ff}=0$.\cite{wang06}


Numerical work on randomness in the AF-singlet transition
has thus far focussed mainly on the strong coupling, Heisenberg spin
limit.  For example, Sandvik has studied how the removal of pairs of
sites in a Heisenberg bilayer affects the AF-singlet quantum critical
point (QCP) in the uniform system.\cite{sandvik06b} 
In addition to the effect of impurities on the AF-singlet transition in
spin models, much is known concerning the simpler problem of the effect
of a single impurity in an AF, both from numerical
and field theoretical work\cite{foot1}.
A focus of much of this past work has been on the behavior of the
uniform susceptibility at the impurity site, where a leading order
$\chi_{\rm imp} \sim 1/T$ Curie divergence is predicted, as well as a
subleading logarithmic divergence.  A careful QMC study of different
types of impurities yields considerable insight into the origins of the
various contributions to $\chi_{\rm imp}$ \cite{hoglund04}.

The distortion of the AF `spin texture' in the neighborhood of a
'dangling impurity' has also been explored\cite{hoglund07}.  Such a
situation arises, for example, when a spin in one layer of a Heisenberg
bilayer is removed, leaving an uncompensated spin-1/2.  In this case,
certain universal physics is predicted to occur, including power law
decays of the spin correlations.  Accurate determinations of the
associated exponents are available \cite{sachdev99,hoglund07}.

In this paper, we
use Exact Diagonalization (ED) and the Determinant Quantum Monte Carlo
(DQMC) methods to examine the physics of a single impurity in the PAM.
Specifically, we compute the suppression of the singlet correlations
about an impurity whose hybrization is in the AF regime, and which is
embedded in a bulk singlet phase.  By computing the AF correlations we
can also infer the size of the AF region about the impurity.  We examine
the implications of these calculations on experiments on disordered
heavy fermion materials.

\section{Model and Methods}

The uniform PAM describes a non-interacting conduction (d) band hybridized
with localized (f) electrons,
\begin{eqnarray}
    {\cal H} = &-&t \sum\limits_{\langle ij \rangle,\sigma}
(d^{\dagger}_{i\sigma}d_{j\sigma}^{\vphantom{dagger}}
+d^{\dagger}_{j\sigma}d_{i\sigma}^{\vphantom{dagger}})
        -V \sum\limits_{i\sigma}
(d^{\dagger}_{i\sigma}f_{i\sigma}^{\vphantom{dagger}}+
f^{\dagger}_{i\sigma}d_{i\sigma}^{\vphantom{dagger}})
\nonumber \\
        &+& U_{\rm f} \sum\limits_{i} (n^{\rm f}_{i\uparrow}-\frac{1}{2})
(n^{\rm f}_{i\downarrow}-\frac{1}{2})
\label{eq:PAM}
\end{eqnarray}
Here $t$ is the hybridization between conduction orbitals with
creation(destruction) operators $d_{i \sigma}^{\dagger} (d_{i
\sigma}^{\phantom{\dagger}})$ on near neighbor sites $\langle ij
\rangle$.  $U_{\rm f}$ is the on-site interaction between spin up and
spin down electrons in a collection of localized orbitals with
creation(destruction) operators $f_{i \sigma}^{\dagger} (f_{i
\sigma}^{\phantom{\dagger}})$ and number operators $n^{\rm
f}_{i\sigma}$.  $V$ is the conduction-localized orbital hybridization.
We have written the interaction term in $H$ in `particle-hole' symmetric
form, so that the lattice is half-filled for all temperatures $T$ and
Hamiltonian parameters $t,U_{\rm f},V$.  Half-filling optimizes the
tendency for AF correlations, and also allows DQMC simulations at low
temperature, since the sign problem is absent.\cite{loh90}
Our investigations here will be on a modification of 
Eq.~\ref{eq:PAM} in which we introduce an ``impurity"
by changing the fd hybridization $V$ to $V_*$ on a
single site $j_0$ of the lattice.

The magnetic physics of the PAM can be characterized by the spin-spin
correlations, between two local ($f$) orbitals and between a local and a
conduction ($d$) orbital:
\begin{align}
\tilde c^{\phantom{f}}_{\rm ff}(j+r,j) &=
\langle \, (n^{\rm f}_{j+r \uparrow} -n^{\rm f}_{j+r \downarrow})
(n^{\rm f}_{j \uparrow} -n^{\rm f}_{j \downarrow}) \, \rangle
\nonumber
\\
\tilde c^{\phantom{f}}_{\rm fd}(j+r,j) &=
\langle \, (n^{\rm f}_{j+r \uparrow} -n^{\rm f}_{j+r \downarrow})
(n^{\rm d}_{j \uparrow} -n^{\rm d}_{j \downarrow}) \, \rangle
\label{spincorr}
\end{align}
These are translationally invariant, that is, functions only of
separation $r$, for the uniform model and periodic boundary conditions.
$\tilde c_{\rm ff}(j+r,j)$ characterizes the range of the intersite
magnetic correlations between the local moments (mediated by the
conduction electrons).  One often focuses primarily on the local, $r=0$
value $\tilde c_{\rm fd}(j,j)$ since it measures the on-site singlet
correlations between the local and conduction electrons.

However, when an impurity is present at site $j_0$, 
translation invariance is broken.
Since we are primarily interested in the alteration of magnetic order in
the vicinity of the impurity, we will focus on the quantities
\begin{align}
c^{\phantom{f}}_{\rm ff}(r) &=
\tilde c^{\phantom{f}}_{\rm ff}(j_0+r,j_0) 
\nonumber \\
c^{\phantom{f}}_{\rm fd}(r) &=
\tilde c^{\phantom{f}}_{\rm fd}(j_0+r,j_0+r) 
\label{spincorr2}
\end{align}
Note that the meaning of $r$ is somewhat different for these two
quantities.  In the case of $c_{\rm ff}(r)$, the distance $r$ represents
how far a second $f$ moment is from the $f$ moment on the impurity site
$j_0$ and thus has the `usual' meaning as a separation of two
spin variables.  $c_{\rm ff}(r)$ measures the range of AF
correlations as one moves away from the impurity.  In the case of
$c_{\rm fd}(r)$, both spins are at a {\it common} distance $r$ from the
impurity-  they sit on the same site- and $r$ represents the distance of
this pair of spins from the impurity.  This is a useful definition, as
we shall see, because it measures the `hole' which is created in a
background of well-formed singlets which is created by a reduction of
$V$ at site $j_0$.

One way to quantify the effect of the impurity is through
the changes in $c_{\rm ff}$ and $c_{\rm fd}$ due to its introduction.
We define the impurity susceptibilities,
\begin{align}
\chi^{\rm imp}_{\rm ff}(r) = \frac{d c_{\rm ff}(r)}{dV}\Big|_{V_*=V}
\label{spinsusc}
\end{align}
to examine how the ff spin correlations between sites $j_0$ and
$j_0+r$ change due to a shift in impurity hybridization $V_*$.

The PAM can be solved exactly on small lattices using the Lanczos
algorithm.  Early work was on two and four site clusters in one
dimension and found that small half-filled lattices are in a singlet
phase for all $U_{\rm f}$ with, however, a near instability to AF.
\cite{jullian82} This work was extended to larger lattices, in one
dimension, by DQMC.\cite{blankenbecler87} Here we will use ED to examine
a single impurity in a one dimensional PAM to gain initial insight into
the effects on the AF and singlet correlations.  The impurity will be
characterized by a reduced value of the $f$-$d$ hybridization $V_*$.  A
focus will be on examining how the bulk $f$-$d$ hybridization affects
the changes induced by the impurity.  We will study geometries in which
the impurity is placed at the end of the one-$d$
chain.  In both cases open boundary conditions are used.  

\begin{figure}
\includegraphics[width=8.0cm]{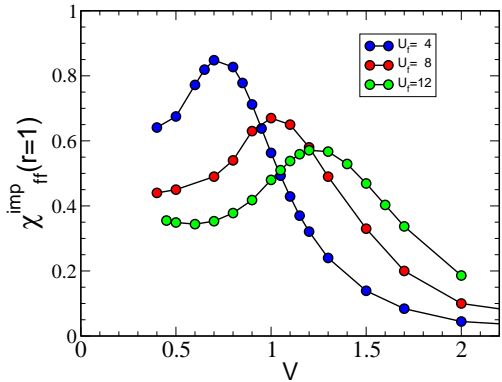}
\caption{(Color online)
Exact diagonalization results for the f orbital impurity
susceptibility on near neighbor sites.  $\chi^{\rm imp}_{\rm ff}(r=1)$
probes the response of AF correlations to a local shift in $V
\rightarrow V_*=V-dV$.  $\chi^{\rm imp}_{\rm ff}(r=1)$ is peaked at
intermediate values of fd hybridization: At large $V$, deep in the
singlet phase, $\chi^{\rm imp}_{\rm ff}(r=1)$, is small.  Similarly,
$\chi^{\rm imp}_{\rm ff}(r=1)$ also declines as $V \rightarrow 0$, deep
in the AF regime.  A maximum occurs at intermediate $V$.  Results are
given for onsite f electron repulsion $U_{\rm f}=4,8,12$.
\label{CR2Vp-V01}
}
\end{figure}

ED, while providing useful initial insight, can examine only small
numbers of sites; hence the focus on the 1D PAM.  Results on larger
lattices can be obtained with the Determinant Quantum Monte Carlo
method,\cite{blankenbecler81} which we will use to study the PAM on a
square lattice (i.e.~in two dimensions).  The DQMC approach introduces a
space and imaginary time dependent classical field to decouple the
interaction $U_{\rm f}$, allowing the fermion degrees of freedom to be
integrated out analytically.  The Boltzmann weight of the resulting
classical Monte Carlo involves determinants of matrices (one for spin up
and one for spin down) of dimension the number of sites in the lattice
$N$.  The computational cost scales as $N^3$, allowing for simulations
on lattices at least an order of magnitude larger than for ED, where the
cost scales exponentially with $N$.  A limitation of DQMC is the `sign
problem' which occurs when the fermion determinants become
negative.\cite{loh90}  The sign problem does not occur in the
Hamiltonian studied here, owing to its particle-hole symmetry (even in
the presence of impurities), which guarantees that the spin up and spin
down determinants have the same sign, so their product is always
positive.

In this project,
we have used ``QUEST," a version of DQMC 
which allows the easy implementation of general geometries
such as the impurity problem considered here.
QUEST also contains a number of modifications to
our ``legacy" codes which improve speed and numerical
stability.  Some of its features are described in 
\cite{bai09,lee10,tomas12,gogolenko14,chang15}.

\section{Results- Exact Diagonalization}

We begin our analysis of the effects of impurities in the PAM with exact
diagonalization (Lanzcos).  While these are on small lattices, they have
the advantage of easily accessing the ground state and large
$U_{\rm f}$, both of which are more challenging in DQMC.  Because of the
smallness of our cluster, $N=6$ sites, we use open boundary conditions.
The impurity is placed at one end of the chain.

We begin by considering the effect of a reduced hybridization $V_*$ on
the spin correlation between near-neighbor f orbitals.  Figure
\ref{CR2Vp-V01} shows the impurity susceptibility on the localized
orbitals, defined in Eq.~\ref{spinsusc}.  $\chi^{\rm imp}_{\rm ff}(r=1)$
is peaked at intermediate values of the bulk $V$, i.e.~between the AF
and singlet phases.  Although there is no AF-singlet transition in
$d=1$,\cite{jullian82} the maximum in $\chi^{\rm imp}_{\rm ff}(r=1)$ is
at crossover values roughly corresponding to the $d=2$ transition and
also shifts to larger $V$ as $U_{\rm ff}$ grows, as previously observed
in DQMC calculations.\cite{vekic95}

\begin{figure}
\includegraphics[width=\linewidth]{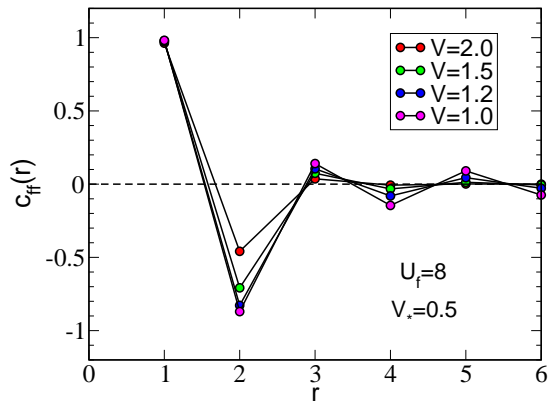}
\caption{(Color online)
ff spin correlation function $c_{\rm ff}(r)$ for an impurity with
$V_*=0.5$ and different bulk $V$.  As $V$ approaches the singlet-AF
crossover, spin correlations at long range develop.  The f-orbital
on-site repulsion $U_{\rm f}=8$.
\label{cffUf8V1p2}
}
\end{figure}

\begin{figure}
\includegraphics[width=\linewidth]{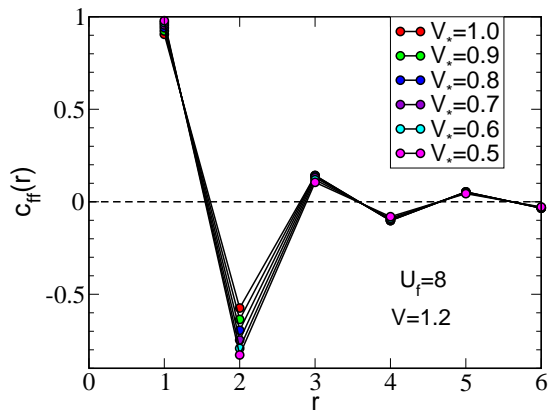}
\caption{(Color online)
Spin correlations in a system with bulk $V=1.2$ for different impurity
$V_*$.  While there is some increase in the near-neighbor ff spin
correlation function $c_{\rm ff}(r=1)$, values at larger separation are
quite insensitive to $V_*$.  Here the f-orbital
on-site repulsion $U_{\rm f}=8$.
\label{cffUf8Vp0p5}
}
\end{figure}

The AF correlations around an impurity site are shown in
Figs.~\ref{cffUf8V1p2} and \ref{cffUf8Vp0p5}.  Figure \ref{cffUf8V1p2}
contains the ff spin correlation function $c_{\rm ff}(r)$ at fixed
impurity hybridization $V_*=0.5$ and varying bulk hybridization $V$.
There is an almost perfectly formed local moment $c_{\rm ff}(r=0)
\approx 1$  and strong near-neighbor spin correlations $c_{\rm
ff}(r=1)$, which grow as $V$ is reduced towards the location of the
AF-singlet crossover.  For $V=2$, the AF correlations are short-ranged,
i.e.~$c_{\rm ff}(r>1) \approx 0$.  But an AF ``droplet" develops around
the impurity as $V$ is decreased, so that by the time $V=1.0$ there is a
clearly discernable correlation even at $r=6$, the maximal separation
accessible on our cluster.

Figure \ref{cffUf8Vp0p5} complements Fig.~\ref{cffUf8V1p2} by providing
data for fixed $V$ and varying $V_*$.  While the near neighbor
correlation $c_{\rm ff}(r=1)$ shows some sensitivity to $V_*$, the
longer range correlations $c_{\rm ff}(r>1)$ are unchanged as $V_*$
varies.  The conclusion of these data is that the AF droplet is quite
sensitive to the bulk hybridization, and develops to quite large
correlation length $\xi$ near the AF-singlet cross-over, but that the
amount of reduction of the impurity hybridization $V_*$ from the bulk
value has little effect on $\xi$.

\section{Results- Quantum Monte Carlo}

We now turn to results obtained with DQMC on $N=$8x8 lattices, much
larger than the $N=6$ cluster studied in exact diagonalization (ED).  We
used open boundary conditions on the small ED lattice, with the impurity
at one end of the chain, to enable the study of effects of the impurity
at distances $r$ reasonably far away.  The 8x8 lattice accessible in
DQMC is large enough to allow data up to $r=4 \sqrt{2}$, even with the
use of periodic boundary conditions (pbc).  This of course has the
advantage of eliminating edge effects, so that the only breaking of
translation invariance is due to the impurity.  In the remainder of this
section we first focus on the effect of the impurity on the spin-spin
correlation function, and then use this data to infer trends in the
correlation length associated with the defect.

It is worth emphasizing that getting to large enough $\beta$ (low $T$)
poses some challenges for the DQMC simulation:  The raw simulation
time naively grows as $\beta$, but more realistically as 
$\beta^p$ where $p \approx 2$ since there is an increase in statistical
noise (requiring longer runs to get the same error bar)
and also an accumulation of round off errors which necessitates more
frequent re-orthogonalization of the matrix products.

\subsection{Spatial variation of singlet correlations
in the vicinity of the impurity}

\begin{figure}
\includegraphics[width=\linewidth]{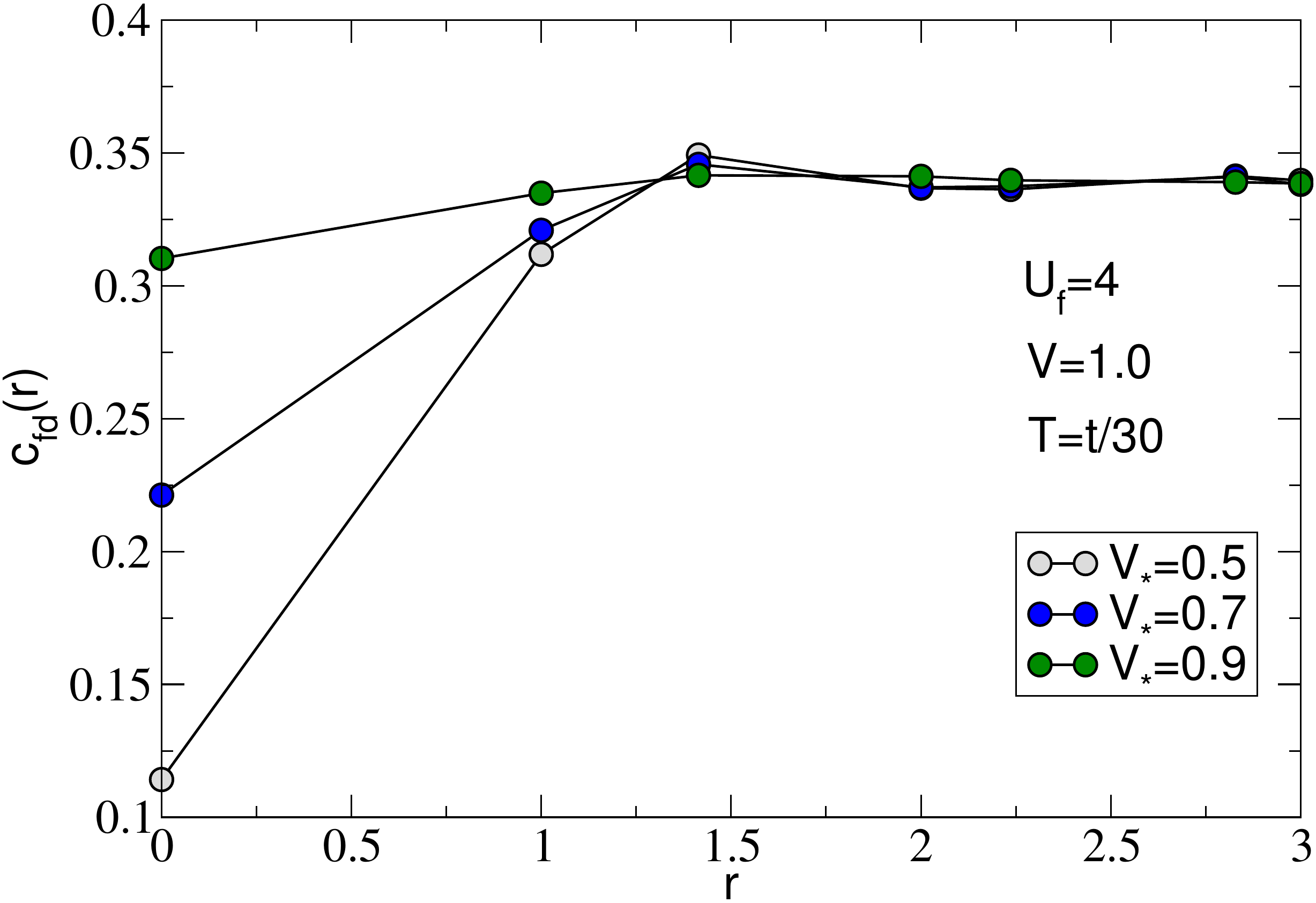}
\caption{(Color online) 
The $fd$ singlet correlator $c_{\rm fd}(r)$ as a function of distance $r$ from the
impurity is shown for bulk hybridization $V=1$ and different impurity
hybridizations $V_*$.  Moving the impurity deeper into the AF phase
steadily reduces the singlet directly on the defect site ($r=0$).
However, the spatial extent of this `hole' does not
increase as $V_*$ decreases.  Here the bulk hybridization $V=1.0$.  The
temperature $T=t/30$. The f-orbital on-site repulsion $U_{\rm f}=4$.
\label{c_cf-vs-r-V10B30}
}
\end{figure}

One way to characterize the effect of a magnetic impurity is to consider
the size and range of the `hole' it digs in the bulk singlet
corrrelations.  In Fig.~\ref{c_cf-vs-r-V10B30} the effect of changing
the strength of the impurity $V_*=0.1, \,0.3, \,0.5$ on the local singlet
correlator $c_{\rm fd}(r)$ 
at distance $r$ from the defect is examined.  Here we choose
$V=1.0$ which is near the bulk AF-singlet boundary.\cite{vekic95}  There is a
systematic reduction in the singlet $c_{\rm fd}(r=0)$ directly on the
defect as its coupling $V_*$ moves deeper into the AF regime.  The range
of the reduction of singlets on $r \neq 0$ sites in its vicinity is
basically independent of $V_*$.   Indeed, $c_{\rm fd}(r)$ does not
differ appreciably from the bulk $V_*=V$ values beyond on-site ($r=0$)
and near-neighbor ($r=1$) separations.  


\begin{figure}
\includegraphics[width=\linewidth]{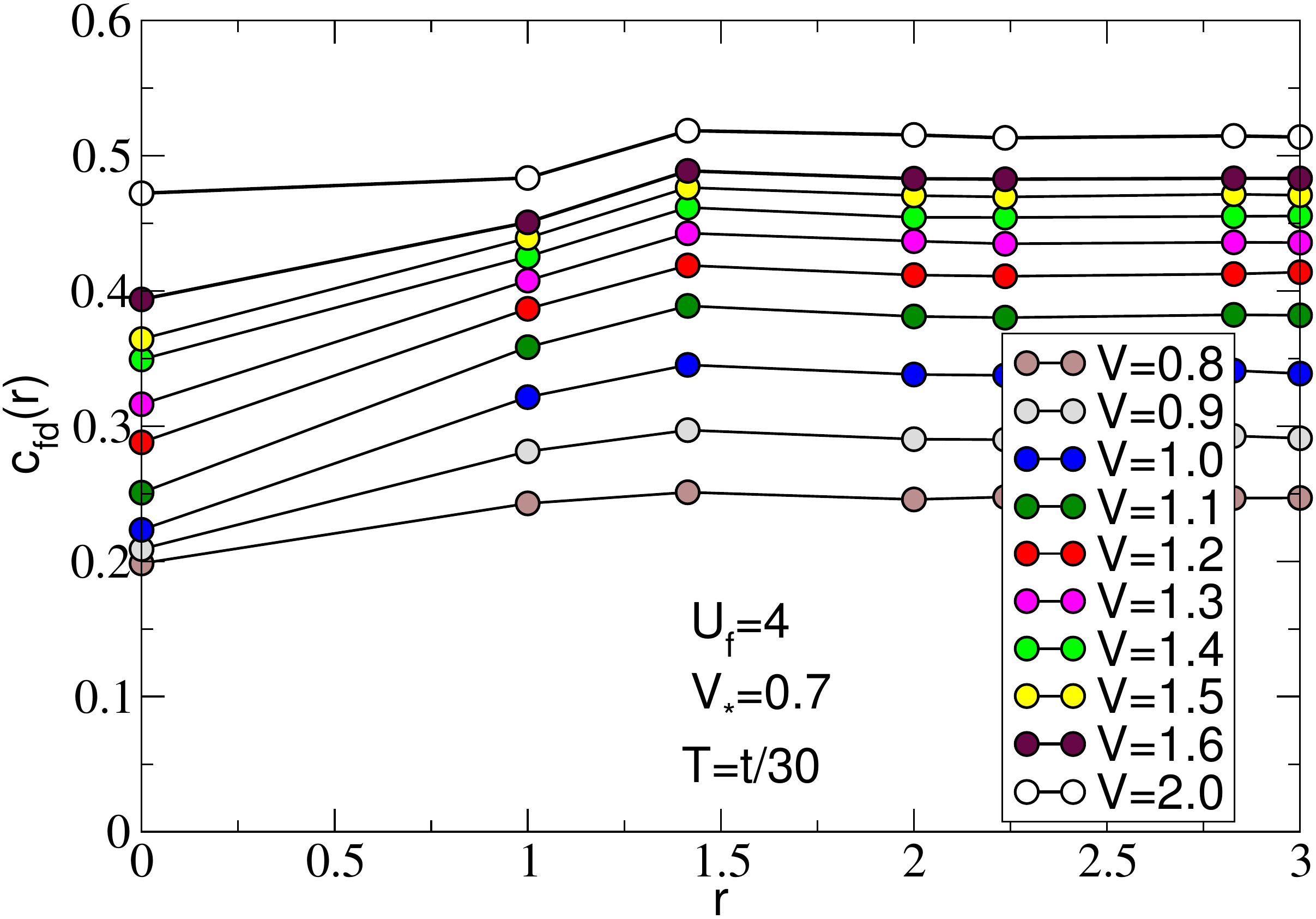}
\caption{(Color online) 
Spin correlations between local and conduction electrons, $c_{\rm
fd}(r)$, are shown for different bulk $V$ and fixed impurity $V_*=0.7$.
All curves exhibit a similar short-range reduction of singlet
correlations near the impurity. However this reduction is largest for
intermediate $V$.   (See also Fig.~6.) The temperature $T=t/30$.  The
f-orbital on-site repulsion $U_{\rm f}=4$.
\label{c_cf-vs-r-Vp07B30}
}
\end{figure}

Figure \ref{c_cf-vs-r-Vp07B30} complements Fig.~\ref{c_cf-vs-r-V10B30}
by presenting the local-conduction spin correlations, $c_{\rm fd}(r)$,
for fixed impurity $V_*=0.7$ and different bulk $V$.  There is a uniform
upward shift for all $r$ in the curves with larger $V$, reflecting the
greater tendency for singlet formation throughout the lattice.  The
suppression of the singlet correlations at the position of the defect
$r=0$ confirm the non-monotonic trend of Fig.~\ref{CR2Vp-V01}.

\begin{figure}
\includegraphics[width=\linewidth]{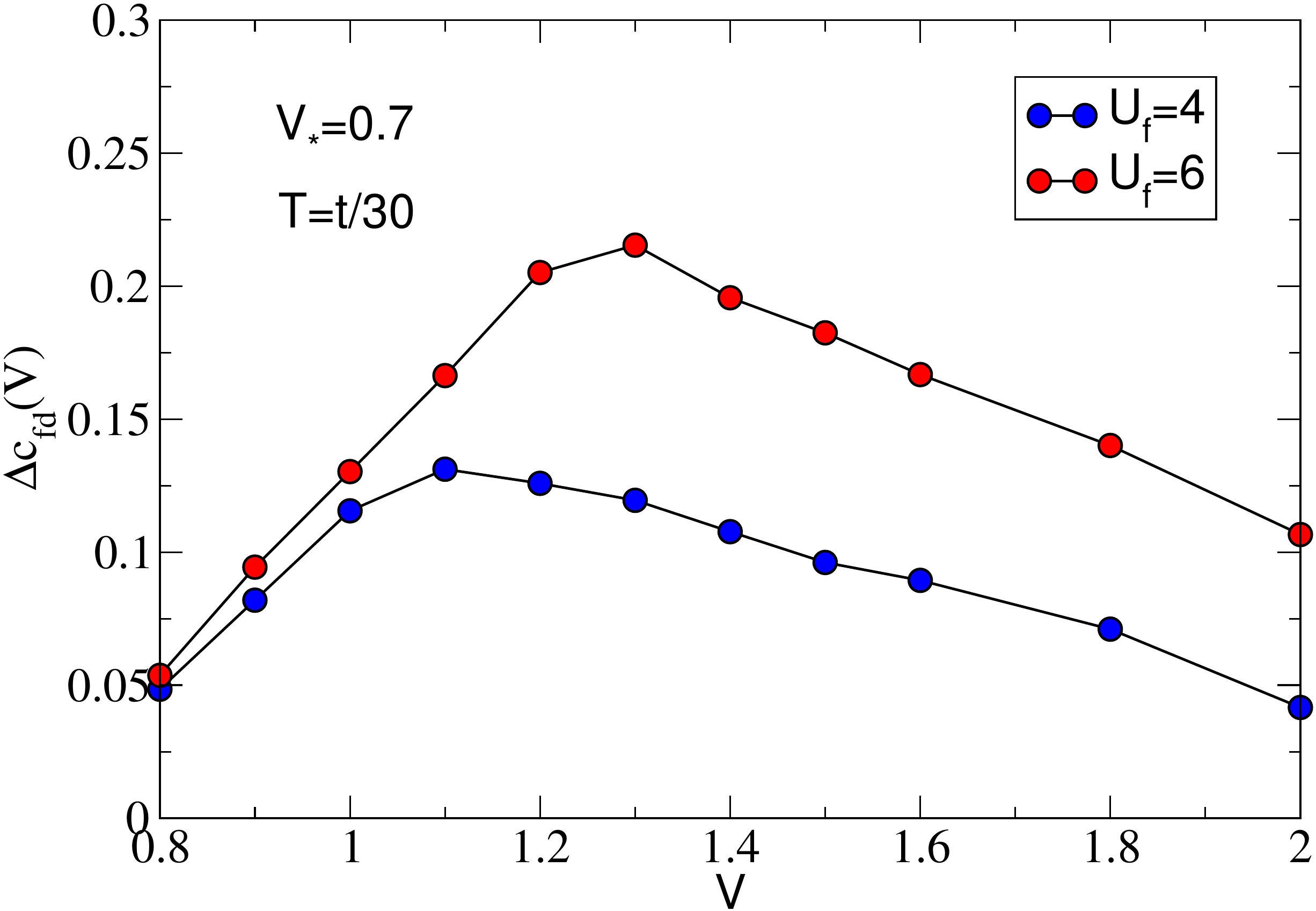}
\caption{(Color online) 
The reduction of the local-conduction spin correlator, $c_{\rm
fd}(r=0)$, from the asymptotic value $c_{\rm fd}(r\rightarrow \infty)$
shown for different bulk hybridization $V$.  The temperature $T=t/30$.
Results are shown for on-site f electron repulsion $U_{\rm f}=4,6$.
Here, for $U_{\rm f}=4$, the 'hole' gets deeper as $V$ is reduced from
$V=2.0$ and becomes deepest for $V=1.1$, after which it is again reduced. 
This tendency fits well with the picture that the effect of an impurity is
largest near the AF-singlet QCP, eg as we show in 
Fig.~\ref{CR2Vp-V01}.  
\label{c_infinity-c_0_Vp07_beta30}
}
\end{figure}

This is more cleanly presented in
Fig.~\ref{c_infinity-c_0_Vp07_beta30}.  We calculate the reduction of
$fd$ correlations in the vicinity of the defect, \begin{eqnarray} \Delta
c_{\rm fd}=c_{\rm fd}(r \rightarrow \infty)-c_{\rm fd}(r=0)
\label{deltac} \end{eqnarray} Fig.~\ref{c_infinity-c_0_Vp07_beta30}
shows the bulk hybridization dependence of $\Delta c_{\rm fd}(V)$. The
largest reduction occurs at intermediate $V$, ie in the vicinity of the
AF-singlet quantum phase transition, in accordance with the ED results
shown in Fig.~\ref{CR2Vp-V01}.  We also calculate $\Delta c_{\rm
fd}(V)$ at higher f-orbital on-site repulsion $U_{\rm f}=6$.  The
occurrence of an `optimal $V_*$' is even more evident, and, to the
extent that the position of this maximum acts as a proxy for the
location $V_c$ of the AF-singlet transition in the uniform periodic
Anderson model, indicates that $V_c$ is an increasing function of $U_f$.
This is consistent with the phase diagram of \cite{vekic95}

\begin{figure}
\includegraphics[width=\linewidth]{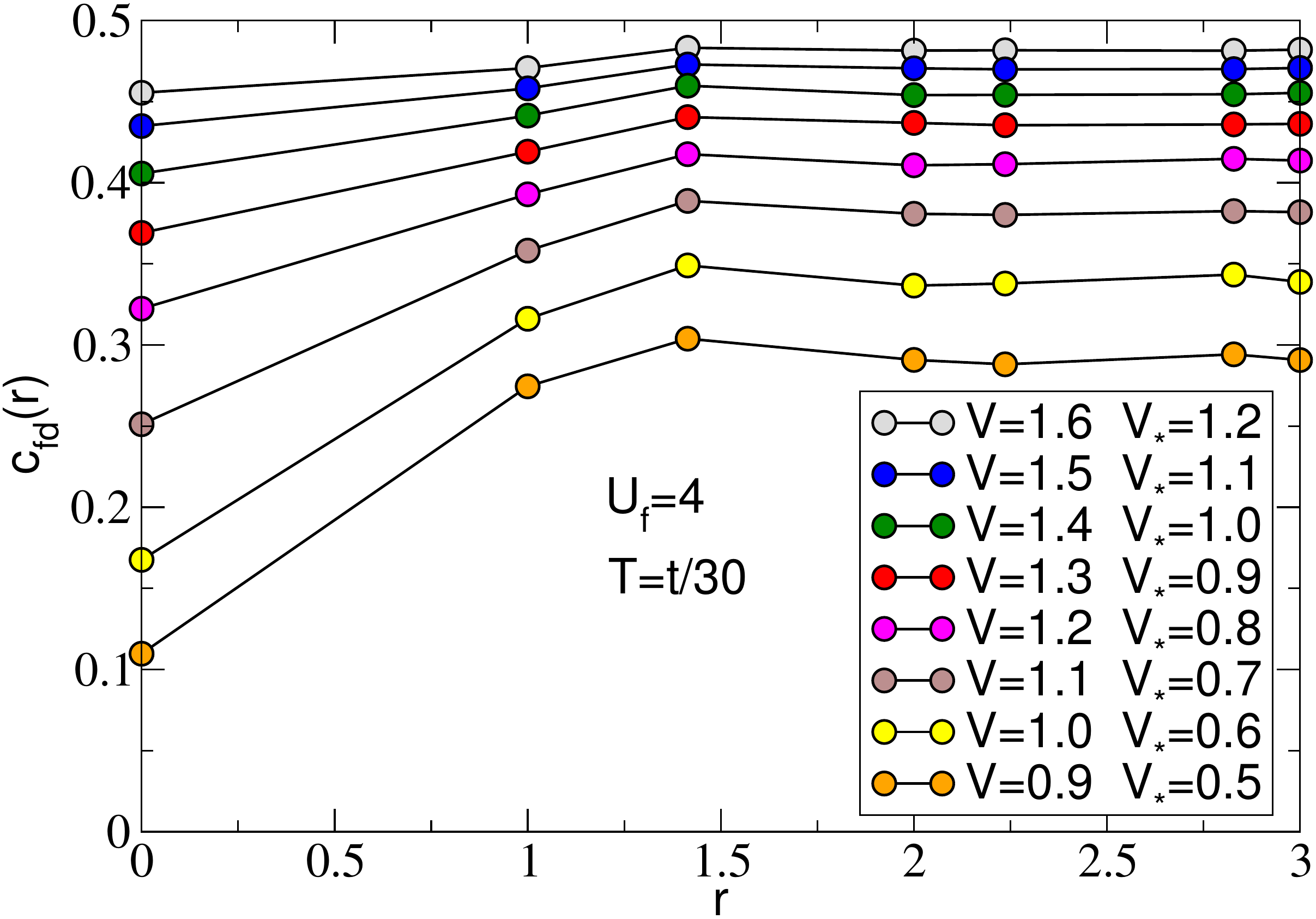}
\caption{(Color online) 
On-site singlet correlator $c_{\rm fd}(r)$ as a function of distance $r$
from the defect.  Here the bulk hybridization $V$ is varied at {\it
fixed} $V-V_*=0.4$.  The maximum in the reduction $\Delta c_{\rm fd}(V)$
at intermediate $V$ is absent.  The temperature $T=t/30$ and the
$f$-orbital on-site repulsion $U_{\rm f}=4$.
\label{c_cf-vs-r-V-Vp04B30}
}
\end{figure}

Because $V_*$ is fixed in
Figs.~\ref{c_cf-vs-r-Vp07B30},\ref{c_infinity-c_0_Vp07_beta30}, as $V$
increases the `defect strength' $V-V_*$ is also increasing.  It is also
interesting to ask what happens as a function of bulk $V$ if $V-V_*$ is
held fixed, Fig.~\ref{c_cf-vs-r-V-Vp04B30}.  Here one sees a monotonic
behavior of $\Delta c_{\rm fd}(V)$.  

\subsection{Spatial variation of AF correlations
in the vicinity of the impurity}

After characterizing the effect of the AF impurity in terms of its
effect on the on-site singlet correlation, we turn now to a
deterimination of the size and strength of the AF `droplet' it induces.

\begin{figure}
\includegraphics[width=\linewidth]{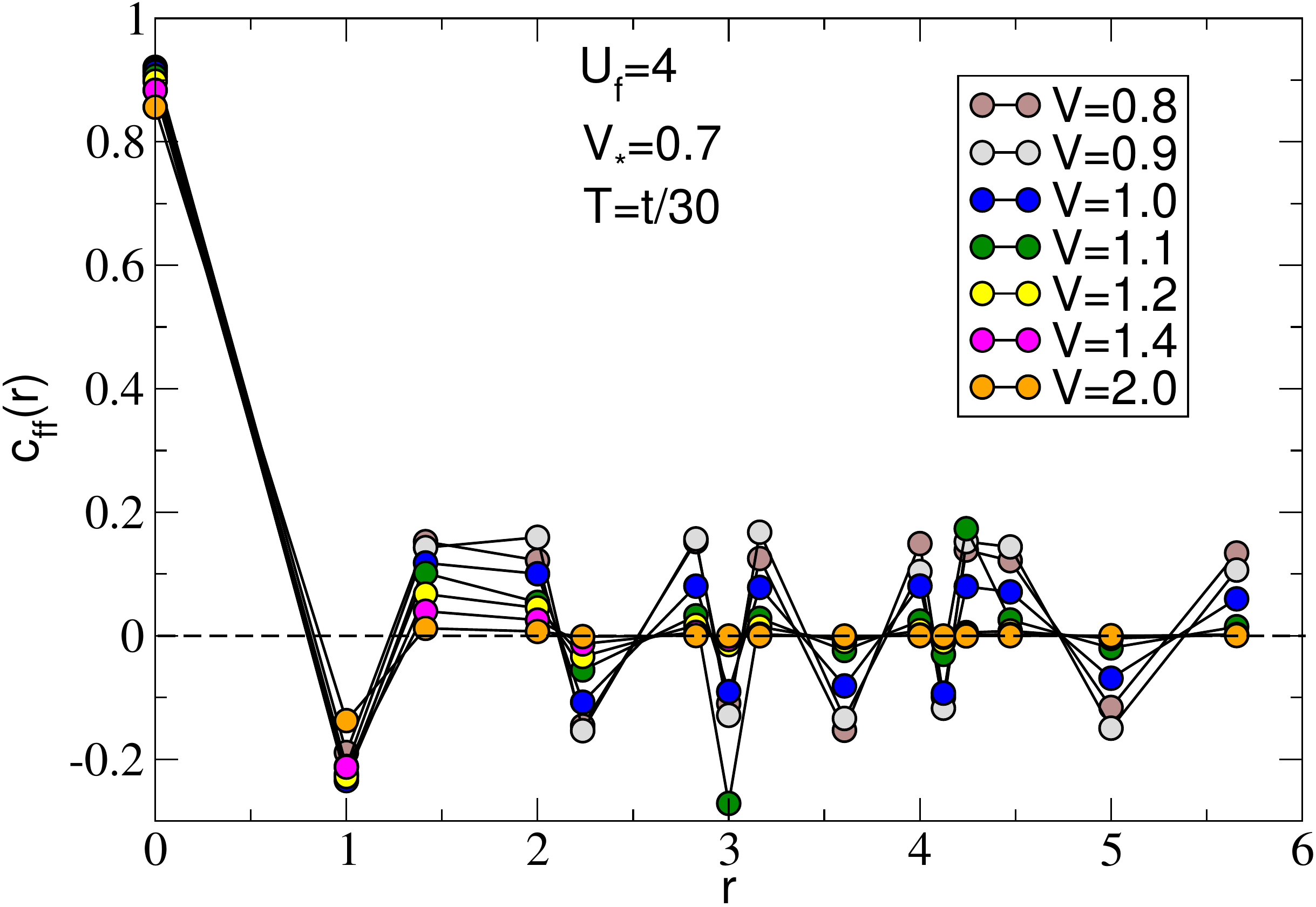}
\caption{(Color online)  
Spin correlation function $c_{\rm ff}(r)$ between two local ($f$)
orbitals.  A defect with impurity hybridization $V_*=0.7$ is present in
different background materials with varying $V$.  There are long range
correlations for $V$ in the AF regime, but the impurity can only induce
AF order locally if $V$ is large.  The temperature $T=t/30$  and the
f-orbital on-site repulsion $U_{\rm f}=4$.
\label{c-vs-r-Vp07B30}
}
\end{figure}

\begin{figure}
\includegraphics[width=\linewidth]{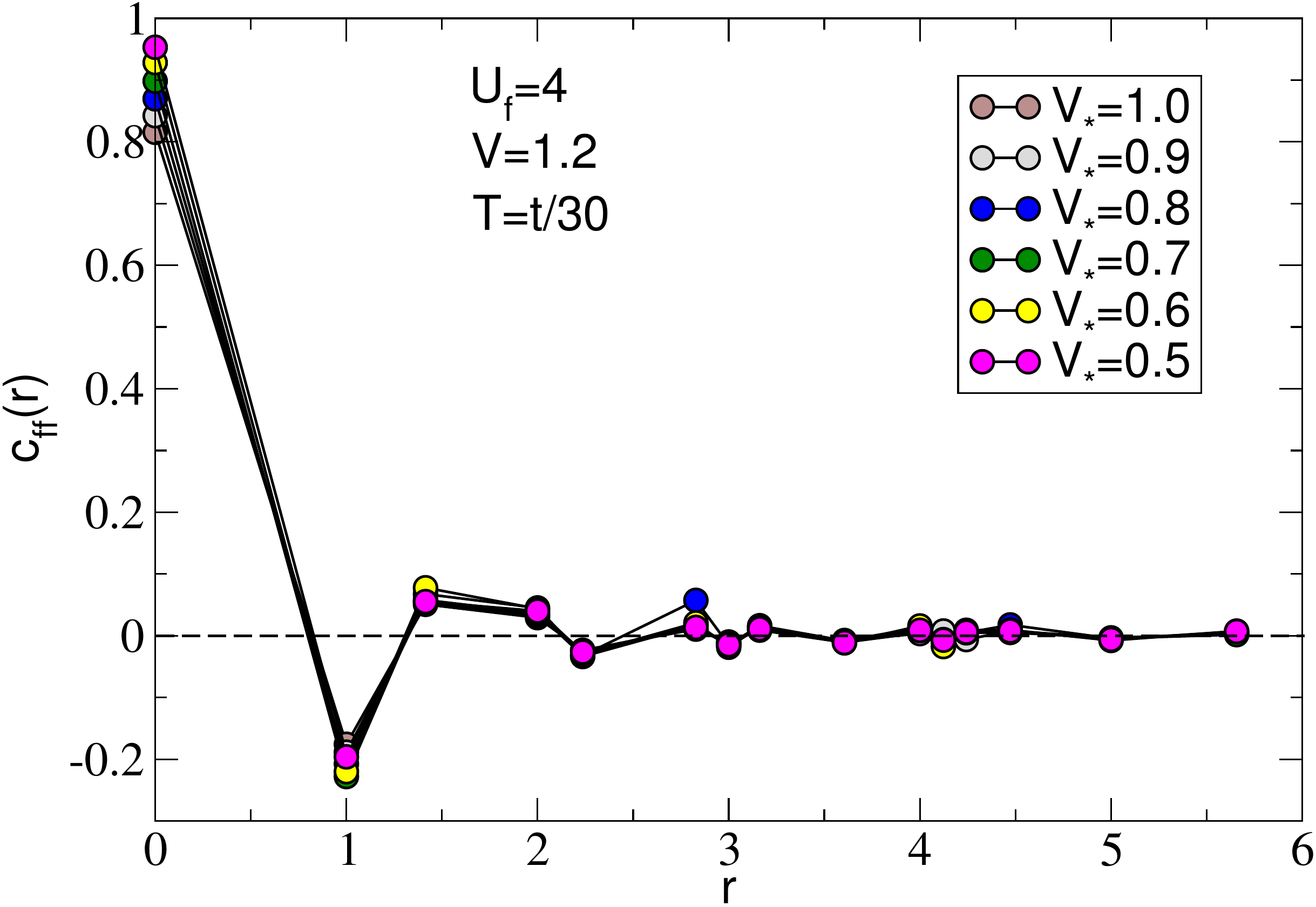}
\caption{(Color online) 
ff electron spin correlation function $c_{\rm ff}(r)$ for bulk
hybridization $V=1.2$ and different impurity hybridizations $V_*$. As
$V_*$ decreases the short-range AF order with near neighbor $(1,0)$
becomes deeper.  The
temperature $T=t/30$. The f-orbital on-site repulsion $U_{\rm f}=4$.
\label{c-vs-r-V12B30}
}
\end{figure}

We begin, in Fig.~\ref{c-vs-r-Vp07B30}, by showing the spin-spin
correlator $c_{\rm ff}(r)$ between the $f$ moment on the defect site
($V_*=0.7$) and an $f$ moment at separation $r$.  (See
Eq.~\ref{spincorr2}.) When the bulk $V=0.8, \,0.9, \,1.0, \,1.1$ is also
in the AF regime, these spin correlations are long ranged: Long range AF
correlations can only survive when the bulk $V$ is in the AF phase. 
Otherwise the impurity can only create a local AF cloud.
Figure \ref{c-vs-r-V12B30} focuses on these effects in the vicinity of
the AF-singlet transition.  The bulk $V=1.2$ is fixed and different
impurity $V_*$ are considered.  The AF correlations are largely
independent of $V_*$. However, in Fig.~\ref{c-vs-r-Vp07B30}, as
one crosses over into the bulk singlet, $V=1.2, \,1.6, \,2.0$ the range
of the AF droplet is finite.  The suppression of $c_{\rm ff}(r)$ at low
$V$ is a finite temperature effect:  The strength of the RKKY coupling
goes as $V^2$, so as $V$ decreases one needs to go to lower temperatures
to access the ground state.  Our results here are at fixed $\beta=30$.
If we were to lower $T$ further (increase $\beta$) at $V=0.5$ the AF
correlations would substantially increase.


\subsection{AF cloud size and Correlation Length}

\begin{figure}
\includegraphics[width=\linewidth]{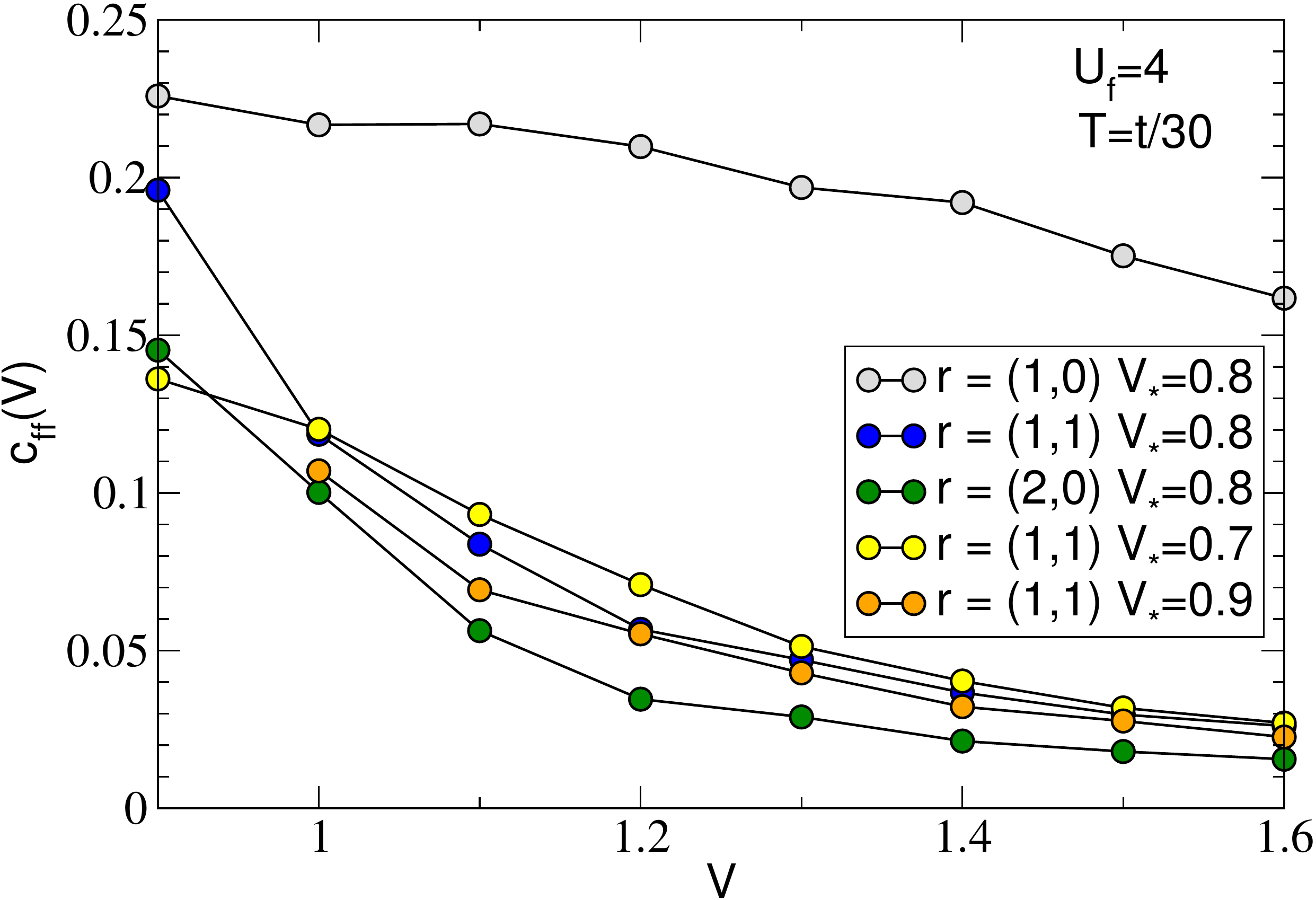}
\caption{(Color online) 
Spin correlations between local electrons, $c_{\rm ff}(r)$, for near
neighbors $(1,0)$, $(1,1)$ and $(2,0)$. $c_{\rm ff}(r)$ grows as $V$
approaches the AF-singlet crossover.  The temperature $T=t/30$. The
f-orbital on-site repulsion $U_{\rm f}=4$.
\label{c-vs-V-r1-r2-Vp08B30}
}
\end{figure}

As discussed in the introduction, a key experimental quantity of
interest is the size of the AF island, e.g.~that created by a Cd
impurity in CeCoIn$_5$.  The preceding figures, which show 
$c_{\rm ff}(r)$, provide a qualitative picture, which here we will
quantify somewhat more precisely.  To begin, it is useful to focus more
closely on the short range spin correlations.  Figure
\ref{c-vs-V-r1-r2-Vp08B30} shows the dependence on $V$ of $c_{\rm
ff}(r)$, for first $r=(1,0)$, second $r=(1,1)$ and third $r=(2,0)$
neighbors.  $c_{\rm ff}(r=1,0)$ is substantial through the AF-singlet
transition region, showing no signature at $V_c$.  This `blindness' to
transitions is of course characteristic of short range correlations.
E.g.~in the Ising model the first neighbor spin
correlation is just the energy, which is smooth through $T_c$.
$c_{\rm ff}(r=1,1)$ and $c_{\rm ff}(r=0,2)$ fall
more rapidly through $V_c$.  Ultimately, of course, at large distances
$c_{\rm ff}(|\vec r| \rightarrow \infty)$ is proportional to the square
of the AF order parameter and would vanish for $V>V_c$.  At $r=(1,1)$,
data for several values $V_*=0.7, \,0.8, \,0.9$ are given.  $c_{\rm
ff}(r=1,1)$ is not very sensitive to the precise value of $V_*$,
especially for $V$ large, where all three curves coincide.
(See Fig.~\ref{c-vs-V-r1-r2-Vp08B30}.)

\begin{figure}
\includegraphics[width=\linewidth]{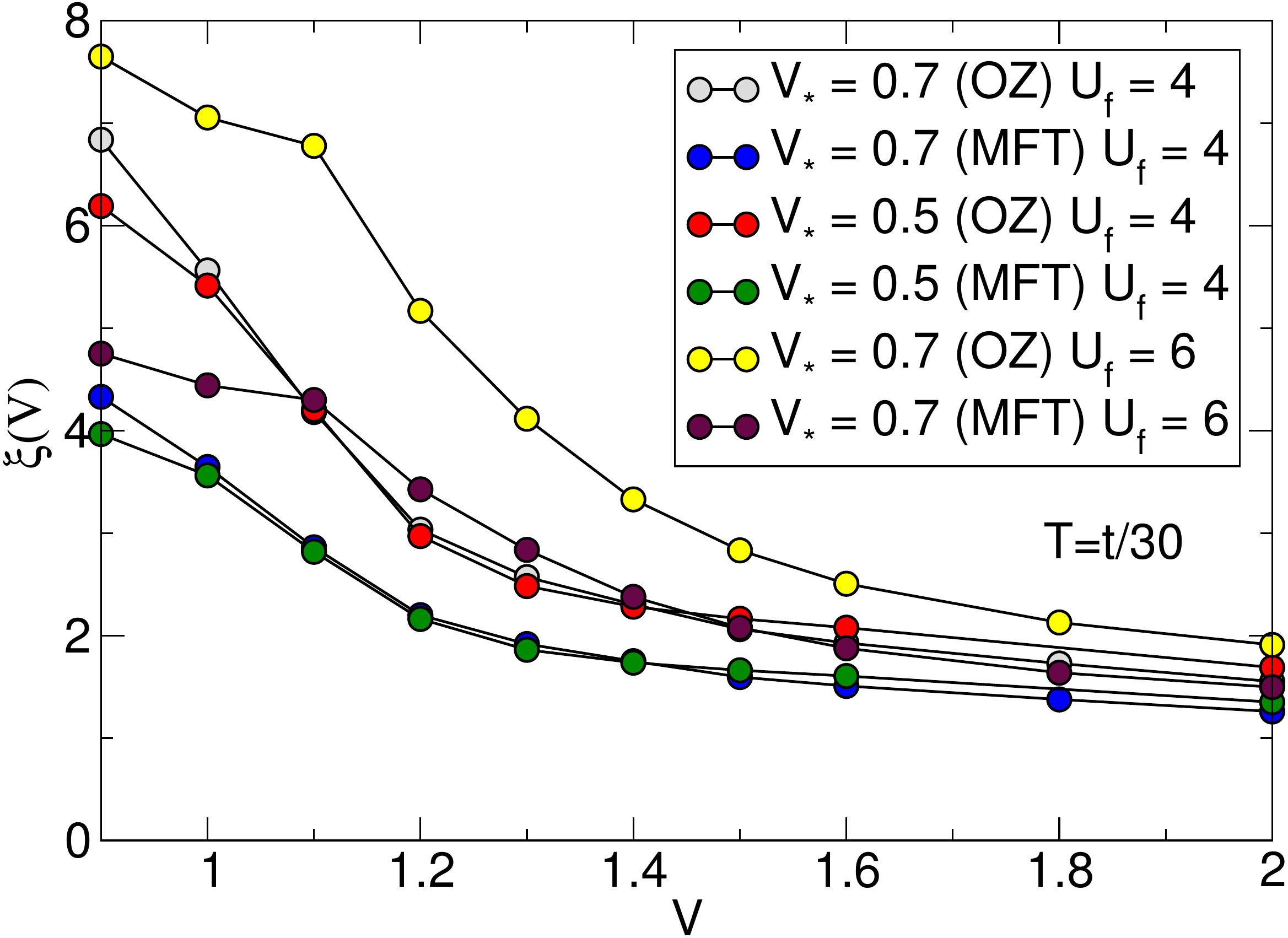}
\caption{(Color online) 
Correlation length versus V given for fixed impurity $V_*=0.5, \,0.7$.
Results are shown for two values of the
on-site f electron repulsion $U_{\rm f}=4,6$. 
OZ (MFT) refer to fits to an Ornstein-Zernike (mean field theory) form.
Although the values of $\xi$ are sensitive to this choice,
the main qualitative feature of the curves is not.  It shows an evolution from 
a flat $\xi \approx 2$ at large $V$ to a rising $\xi \approx 4-6$ 
(at $U_{\rm f}=4$), indicating larger AF regions.
The temperature $T=t/30$.
\label{correlation_length_vs_V_S108_beta30}
}
\end{figure}

The plots of $c_{\rm ff}(r)$ already give a qualitative indication of
the value of the correlation length $\xi$.  We can make this somewhat more
quantitative by fitting the magnitude of $c_{\rm ff}(r)$
either to a mean field or Ornstein-Zernike form to obtain the
AF correlation length $\xi$.  This is not expected to be too precise
since these functional forms describe the asymptotic behavior at large
distances which are not sampled on our finite lattices.  
Nevertheless,
Fig.~\ref{correlation_length_vs_V_S108_beta30} 
shows the results of our DQMC
simulations for the dependence of 
$\xi$ on bulk hybrdization $V$ for fixed impurity $V_*=0.5, \,0.7$
and two values of $U_{\rm f}$.
The conclusion is that the size of the AF cloud
is fairly small, $\xi \lesssim 2$, unless the bulk $V$ is close to, or
below, the AF-singlet QCP.  It takes a value $\xi \sim 3$ right at $V =
V_c \sim 1.1$ \cite{vekic95}. Well into the AF phase, $\xi$ is of
the order of
the linear size of the lattice, indicating that there is long
range order throughout the 8x8 cluster studied here.
Finite size scaling can be used to establish order in the
thermodynamic limit\cite{vekic95,varney09}.
The change in slope from the flat region $\xi \approx 2$
to a rising $\xi$ occurs as one approaches the bulk
critical point $V_c \approx 1.1$.  Indeed, although it is
not so easy to tell from the scatter in the data, there is some
indication of an inflection point (maximal slope of growth
of $\xi$) at the QCP.
It has been noted in quantum spin models\cite{hoglund04,sachdev99}, that
the effect of impurities is localized in the spin gap (singlet) phase,
but that at the QCP critical spin correlations instead exhibit
power law decays in space and imaginary time.

\section{Conclusions}

Much of our understanding, through QMC, of the effects of randomness in
quantum antiferromagnets, and their implications for materials like
disordered heavy fermion compounds has been developed with spin models
like the transverse field Ising or Heisenberg bilayer Hamiltonians.
\cite{sandvik94,sandvik06,sandvik06b,guo96,young97,castroneto98,vojta10}
In this paper, we have examined the physics of a single impurity on the
competition between antiferromagnetic and singlet correlations in the
Periodic Anderson Model, i.e.~in the more computationally challenging
case of an {\it itinerant} electron model.  An impurity with a $d$-$f$
hybridization in the AF regime suppresses singlet correlations in its
vicinity, and at the same time induces an AF domain.  As the $f$-$d$
hybridization of the bulk approaches the AF transition from the singlet
side, the impurity
becomes increasingly effective at inducing AF correlations.

Our work parallels earlier Hartree-Fock studies of disorder in cuprate
superconductors, modeled with the single band Hubbard
Hamiltonian.\cite{andersen07} There, the level of doping $x$ and the
on-site Hubbard interaction $U$ are used to tune the system in the
neighborhood of magnetic and superconducting phase transitions, and the
effect of impurities is then explored.  Analogously, we have here
examined an impurity in a multi-band periodic Anderson model which has a
singlet-AF quantum phase transition.

These results provide an important confirmation for the experimental
observation of AF droplets in quantum critical CeCoIn$_5$.  In this
material, the AF droplets were observed to disappear under pressure.
Increasing pressure corresponds to increasing $V$ as the $d$ and $f$
orbital overlap increases.  As our simulations indicate, the size and
extent of the AF droplets decrease with increasing $V$ as the system is
tuned away from the QCP.  These studies also suggest that heavy fermion
systems that can be tuned to a QCP under pressure, such as CeRhIn$_5$,
may also exhibit similar
behavior.\cite{HeggerRh115discovery,ParkCeRhIn5PhaseDiagram}  An
important question is whether `chemical pressure' in which dopants are
added to either expand or contract the lattice actually affects the
electronic degrees of freedom by modifying the local hybridization.  In
the case of Cd-doped CeCoIn$_5$, the Cd has no 5p electrons to hybridize
with the neighboring Ce, thus our introduction of $V_*<V$ is a
reasonable approach.  Sn-doping, on the other hand, with one extra 5p
electron, does not exhibit AF order.\cite{BauerCeCoIn5SnPRL2005}   

The ED method for the PAM is limited to systems of around ten sites. DQMC
calculations can be performed on several hundred sites.  Even so, in 2D,
the lattices accessible to DQMC are only 10-20 sites in linear extent.
Thus it is difficult to perform definitive studies of multiple
inpurities and their surrounding domains.  This is a crucial question,
because the overlapping of AF domains about individual impurities, and
the lack of frustration in such overlap, is believed to lead to AF long
range order.\cite{kivelson95} Quantum spin models like the Heisenberg
bilayer can, however, be explored on much larger lattices.  Future work
will focus on examining randomness and the AF-singlet transition for
such quantum spin models.

\section{Acknowledgements}

\noindent
We thank T. Park and J.D. Thompson for enlightening discussions.  
Work supported by NNSA DE-NA0002908, and by the Fulbright Foundation.


\bibliographystyle{unsrt}
\bibliography{benali}

\end{document}